\documentclass[aps, pra, twocolumn,showpacs,preprintnumbers,amsmath,amssymb]{revtex4}

\newcommand{\ket}[1]{\mbox{$ | #1 \rangle $}}
\newcommand{\bra}[1]{\mbox{$ \langle #1 | $}}
\usepackage[dvips]{graphicx}

\begin{document}

\preprint{}

\title{Multiparty simultaneous quantum identity authentication based on entanglement swapping}

\author{Jian Wang}

 \email{jwang@nudt.edu.cn}

\affiliation{School of Electronic Science and Engineering,
\\National University of Defense Technology, Changsha, 410073, China }
\author{Quan Zhang}
\affiliation{School of Electronic Science and Engineering,
\\National University of Defense Technology, Changsha, 410073, China }
\author{Chao-jing Tang}
\affiliation{School of Electronic Science and Engineering,
\\National University of Defense Technology, Changsha, 410073, China }


\begin{abstract}
We present a multiparty simultaneous quantum identity authentication
protocol based on entanglement swapping. In our protocol, the
multi-user can be authenticated by a trusted third party
simultaneously.
\end{abstract}

\maketitle


%
%
Quantum cryptography has been one of the most remarkable
applications of quantum mechanics in quantum information science.
Quantum key distribution (QKD), which provides a way of exchanging
a private key with unconditional security, has progressed rapidly
since the first QKD protocol was proposed by Benneett and Brassard
in 1984 \cite{bb84}. A good many of other quantum communication
schemes have also been proposed and pursued, such as quantum
secret sharing (QSS)\cite{hbb99,kki99,zhang,gg03,zlm05,xldp04},
quantum secure direct communication (QSDC)
\cite{beige,Bostrom,Deng,denglong,cai1,cai4,jwang1,jwang2,jwang3,hlee,cw1,cw2,tg,zjz,gao1}
and quantum identity authentication (QIA)
\cite{dhh99,zz00,m02,hlee}. QSS is the generalization of classical
secret sharing to quantum scenario and can share both classical
and quantum messages among sharers. QSDC's object is to transmit
the secret message directly without first establishing a key to
encrypt it. Authentication is a well-studied area of classical
cryptography, including identity and message authentication. QIA
aims to generalize classical identity authentication to quantum
scenario for providing unconditional security. Du$\check{e}$k et
al. \cite{dhh99} proposed a secure quantum identification system
combining a classical identification procedure and quantum key
distribution. Zeng and Zhang \cite{zz00} put forward a quantum key
verification scheme which can simultaneously distribute the
quantum secret key and verify the communicators' identity. T.
Mihara \cite{m02} presented three quantum identification schemes
by using entangled state and unitary operation. Lee et al.
\cite{hlee} presented two QSDC protocols with user authentication.

In this paper, we present a multiparty simultaneous quantum
identity authentication protocol based on entanglement swapping,
which combines the idea in Ref. \cite{gao1} with that in Ref.
\cite{hlee}. In our protocol, We suppose a trusted third party,
Trent, authenticates $r$ legal users, \{Alice$_1$, Alice$_2$,
$\cdots$, Alice$_r$\} simultaneously. Similar to Ref. \cite{hlee},
Trent shares a secret identity number $ID_i$ ($i=1,2,\cdots,r$)
and a secret hash function $h_i$ ($i=1,2,\cdots,r$) with each
user. Here the hash function is defined as
\begin{eqnarray}
h: \{0,1\}^l\times\{0,1\}^m\rightarrow\{0,1\}^n,
\end{eqnarray}
where $l$, $m$ and $n$ denote the length of the identity number,
the length of a counter and the length of authentication key,
respectively. Thus the user's authentication key can be expressed
as $AK=h(ID, C)$, where $C$ is the counter of calls on the user's
hash function. When the length of the authentication key is not
enough to satisfy the requirement of cryptographic task. The
parties can increase the counter and then generates a new
authentication key. We denote the authentication keys of
Alice$_1$, Alice$_2$, $\cdots$, Alice$_r$ as
$AK_{A_1}=h_{A_1}(ID_{A_1}, C_{A_1})$, $AK_{A_2}=h_{A_2}(ID_{A_2},
C_{A_2})$, $\cdots$, $AK_{A_r}=h_{A_r}(ID_{A_r}, C_{A_r})$,
respectively.

Entanglement swapping can entangle two quantum systems that do not
have direct interaction with each other \cite{zzhe93}. It plays an
important role in quantum information. We first describe
entanglement swapping simply. The four Bell states are
\begin{eqnarray}
\ket{\phi^\pm}=\frac{1}{\sqrt{2}}(\ket{00}\pm\ket{11}),\nonumber\\
\ket{\psi^\pm}=\frac{1}{\sqrt{2}}(\ket{01}\pm\ket{10}).
\end{eqnarray}
Suppose two distant parties, Alice and Bob, share \ket{\phi^+_{12}}
and \ket{\phi^+_{34}} where Alice has qubits 1 and 4, and Bob
possesses 2 and 3. Note that
\begin{eqnarray}
\label{1}
\ket{\phi^+_{12}}\otimes\ket{\phi^+_{34}}&=&\frac{1}{2}(\ket{\phi^+_{14}}\ket{\phi^+_{23}}+\ket{\phi^-_{14}}\ket{\phi^-_{23}}\nonumber\\
&
&+\ket{\psi^+_{14}}\ket{\psi^+_{23}}+\ket{\psi^-_{14}}\ket{\psi^-_{23}}.
\end{eqnarray}
After Bell basis measurement on qubits 1 and 4, the state of the
qubits 1, 2, 3, 4 collapses to \ket{\phi^+_{14}}\ket{\phi^+_{23}},
\ket{\phi^-_{14}}\ket{\phi^-_{23}},
\ket{\psi^+_{14}}\ket{\psi^+_{23}} and
\ket{\psi^-_{14}}\ket{\psi^-_{23}} each with probability 1/4. If
Alice and Bob share other Bell states, similar results can be
achieved.

In our protocol, the eight three-particle GHZ states are defined
as
\begin{eqnarray}
\label{1}
\ket{\Psi_1}=\frac{1}{\sqrt{2}}(\ket{000}+\ket{111}), \ket{\Psi_2}=\frac{1}{\sqrt{2}}(\ket{000}-\ket{111}),\nonumber\\
\ket{\Psi_3}=\frac{1}{\sqrt{2}}(\ket{100}+\ket{011}), \ket{\Psi_4}=\frac{1}{\sqrt{2}}(\ket{100}-\ket{011}),\nonumber\\
\ket{\Psi_5}=\frac{1}{\sqrt{2}}(\ket{010}+\ket{101}), \ket{\Psi_6}=\frac{1}{\sqrt{2}}(\ket{010}-\ket{101}),\nonumber\\
\ket{\Psi_7}=\frac{1}{\sqrt{2}}(\ket{110}+\ket{001}), \ket{\Psi_8}=\frac{1}{\sqrt{2}}(\ket{110}-\ket{001}),\nonumber\\
\end{eqnarray}
which form a complete orthonormal basis. The parties agree that
the two unitary operations
\begin{eqnarray}
I=\ket{0}\bra{0}+\ket{1}\bra{1},\nonumber\\
i\sigma_y=\ket{0}\bra{1}-\ket{1}\bra{0},
\end{eqnarray}
can be encoded into one bit classical information as
\begin{eqnarray}
I\rightarrow 0,  i\sigma_y\rightarrow 1.
\end{eqnarray}

We first present our QIA protocol with two users (Alice$_1$,
Alice$_2$) and then generalize it to the case with many users
(Alice$_1$, Alice$_2$, $\cdots$, Alice$_r$). Each user shares a
authentication key with Trent, as we have described above.

(S1) Trent prepares an ordered $N$ three-particle GHZ states, each
of which is in the state
$\ket{\Psi_1}=\frac{1}{\sqrt{2}}(\ket{000}+\ket{111})_{TA_1A_2}$,
where the subscripts $T$, $A_1$ and $A_2$ represent the three
particles of each GHZ state. Trent takes particle $T$ ($A_1$,
$A_2$) for each state to form an ordered particle sequence, called
$T$ ($A_1$, $A_2$) sequence. He then sends $A_1$ and $A_2$
sequences to Alice$_1$ and Alice$_2$, respectively and keeps $T$
sequence.

(S2) To ensure the security of the quantum channel, the parties
check eavesdropping as follows: (a) After hearing from the users,
Trent selects randomly a sufficiently large subset from the
ordered $N$ GHZ states. (b) He measures the sampling particles in
$T$ sequence, in a random measuring basis,
$Z$-basis(\ket{0},\ket{1}) or $X$-basis
(\ket{+}=$\frac{1}{\sqrt{2}}(\ket{0}+\ket{1})$,
\ket{-}=$\frac{1}{\sqrt{2}}(\ket{0}-\ket{1})$). (c) Trent
announces publicly the positions of the sampling particles and the
measuring basis for each of the sampling particles. Alice$_1$
(Alice$_2$) measures the sampling particles in $A_1$ ($A_2$)
sequence, in the same measuring basis as Trent. After
measurements, the users publishes their measurement results. (d)
Trent can then check the existence of eavesdropper by comparing
their measurement results. If the channel is safe, their results
must be completely correlated. When Trent performs $Z$-basis
measurement on his particle, Alices' result should be \ket{00}
(\ket{11}) if Trent's result is \ket{0} (\ket{1}). On the
contrary, Alices' result should be \ket{++} or \ket{--} (\ket{+-}
or \ket{-+}) if Trent performs $X$-basis measurement on his
particle and gets the result \ket{+} (\ket{-}). (e) If Trent
confirms that their results are completely correlated, he
announces publicly his measurement results of the sampling
particles. The users can make certain whether they share a
sequence of GHZ states with Trent. If the users confirms that
there is no eavesdropping, they continue to execute the next step.
Otherwise, they inform Trent and abort the communication.

(S3) After hearing from the users, Trent divides randomly the
remaining GHZ states into $M$ ordered groups, \{P(1)$_{TA_1A_2}$,
Q(1)$_{T'A_1'A_2'}$\}, \{P(2)$_{TA_1A_2}$, Q(2)$_{T'A_1'A_2'}$\},
$\cdots$, \{P(M)$_{TA_1A_2}$, Q(M)$_{T'A_1'A_2'}$\}, where 1, 2,
$\cdots$, $M$ represent the order of the group and the subscripts
$T$ and $T'$ ($A_1$, $A_1'$ and $A_2$, $A_2'$) denote the
particles belonging to Trent (Alice$_1$'s and Alice$_2$'s ).

(S4) For each of the groups, Alice$_1$ (Alice$_2$) performs one of
the two operations \{$I$, $i\sigma_y$\} on particle $A_1$ ($A_2$)
according to her authentication key, $AK_{A_1}$ ($AK_{A_2}$). For
example, if the $i$th value of $AK_{A_1}$ is 0 (1), Alice$_1$
executes $I$ ($i\sigma_y$) operation on particle $A_1$. As we have
described above, here $AK_{A_1}=h(ID_{A_1}, C_{A_1})$,
$AK_{A_2}=h(ID_{A_2}, C_{A_2})$. If the length of $AK$ is not long
enough to $M$, new $AK$ can be generated by increasing the counter
until the length of $AK$ is no less than $M$. They inform Trent
that they have transformed their qubit by using unitary operation
according to their authentication keys.

(S5) After hearing from the users, Trent performs randomly $I$ or
$i\sigma_y$ operation on particles $T$ in each group. After the
three-party's operations, $\ket{\Psi_1}$ can be transformed into
one of the eight three-particle GHZ states \{\ket{\Psi_1},
\ket{\Psi_1}, $\cdots$, \ket{\Psi_8}\}, as shown in Table 1.
\begin{table}[h]
\caption{The transformation relations of GHZ
states}\label{Tab:one}
  \centering
    \begin{tabular}[b]{|c|c|} \hline
       & unitary operations performed on the three paticles\\ \hline
      \ \ket{\Psi_1} & $I\otimes I\otimes I$ \\ \hline
      \ \ket{\Psi_2} & $i\sigma_y\otimes i\sigma_y\otimes i\sigma_y$\\ \hline
      \ \ket{\Psi_3} & $I\otimes i\sigma_y\otimes i\sigma_y$\\ \hline
      \ \ket{\Psi_4} & $i\sigma_y\otimes I\otimes I$\\ \hline
      \ \ket{\Psi_5} & $i\sigma_y\otimes I\otimes i\sigma_y$\\ \hline
      \ \ket{\Psi_6} & $I\otimes i\sigma_y\otimes I$\\ \hline
      \ \ket{\Psi_7} & $i\sigma_y\otimes i\sigma_y\otimes I$\\ \hline
      \ \ket{\Psi_8} & $I\otimes I\otimes i\sigma_y$\\ \hline
    \end{tabular}
\end{table}

(S6) Trent lets Alice$_1$ (Alice$_2$) measure particles $A_1$ and
$A_1'$ ($A_2$ and $A_2'$) of each group in Bell basis. After
measurements, Alice$_1$ and Alice$_1$ publish their measurement
results. Trent performs Bell basis measurement on particles $T$
and $T'$ of each group and authenticates the users according to
their measurement results. We then explain it in detail. The state
of a group can be written as
\begin{eqnarray}
\ket{\Psi_1}_{TA_1A_2}\otimes\ket{\Psi_1}_{T'A_1'A_2'}=
\frac{1}{2\sqrt{2}}(\ket{\phi^+_{TT'}}\ket{\phi^+_{A_1A_1'}}\ket{\phi^+_{A_2A_2'}}\nonumber\\
+\ket{\phi^+_{TT'}}\ket{\phi^-_{A_1A_1'}}\ket{\phi^-_{A_2A_2'}}\nonumber\\
+\ket{\phi^-_{TT'}}\ket{\phi^+_{A_1A_1'}}\ket{\phi^-_{A_2A_2'}}+\ket{\phi^-_{TT'}}\ket{\phi^-_{A_1A_1'}}\ket{\phi^+_{A_2A_2'}}\nonumber\\
+\ket{\psi^+_{TT'}}\ket{\psi^+_{A_1A_1'}}\ket{\psi^+_{A_2A_2'}}+\ket{\psi^+_{TT'}}\ket{\psi^-_{A_1A_1'}}\ket{\psi^-_{A_2A_2'}}\nonumber\\
+\ket{\psi^-_{TT'}}\ket{\psi^+_{A_1A_1'}}\ket{\psi^-_{A_2A_2'}}+\ket{\psi^-_{TT'}}\ket{\psi^-_{A_1A_1'}}\ket{\psi^+_{A_2A_2'}}).\nonumber\\
\end{eqnarray}
If Trent's random operation is $i\sigma_y$, Alice$_1$'s $i$th
value of her authentication key is 1 which corresponds to
operation $i\sigma_y$ and Alice$_2$'s $i$th value of her
authentication key is 0 corresponding to operation $I$,
\ket{\Psi_1}$_{TA_1A_2}$ is then transformed to
\ket{\Psi_7}$_{TA_1A_2}$ and the state of the group becomes
\begin{eqnarray}
\label{es}
\ket{\Psi_7}_{TA_1A_2}\otimes\ket{\Psi_1}_{T'A_1'A_2'}=
\frac{1}{2\sqrt{2}}(\ket{\psi^+_{TT'}}\ket{\psi^+_{A_1A_1'}}\ket{\phi^+_{A_2A_2'}}\nonumber\\
-\ket{\psi^+_{TT'}}\ket{\psi^-_{A_1A_1'}}\ket{\phi^-_{A_2A_2'}}\nonumber\\
-\ket{\psi^-_{TT'}}\ket{\psi^+_{A_1A_1'}}\ket{\phi^-_{A_2A_2'}}+\ket{\psi^-_{TT'}}\ket{\psi^-_{A_1A_1'}}\ket{\phi^+_{A_2A_2'}}\nonumber\\
+\ket{\phi^+_{TT'}}\ket{\phi^+_{A_1A_1'}}\ket{\psi^+_{A_2A_2'}}-\ket{\phi^+_{TT'}}\ket{\phi^-_{A_1A_1'}}\ket{\psi^-_{A_2A_2'}}\nonumber\\
-\ket{\phi^-_{TT'}}\ket{\phi^+_{A_1A_1'}}\ket{\psi^-_{A_2A_2'}}+\ket{\phi^-_{TT'}}\ket{\phi^-_{A_1A_1'}}\ket{\psi^+_{A_2A_2'}}).\nonumber\\
\end{eqnarray}
From the published results of Alice$_1$ and Alice$_2$ and his
measurement results, Trent can obtain the users' operation
information and then authenticates the users because the three
parties' results correspond to an exclusive state. For example,
the results of Trent, Alice$_1$ and Alice$_2$ are each
\ket{\psi^-_{TT'}}, \ket{\psi^-_{A_1A_1'}} and
\ket{\phi^+_{A_2A_2'}}. According to Eq. (\ref{es}), the state of
the group must be
\ket{\Psi_7}$_{TA_1A_2}\otimes$\ket{\Psi_1}$_{T'A_1'A_2'}$. Trent
then knows the $i$th value of Alice$_1$'s and Alice$_2$'s
authentication keys are each 1 and 0 because only the operation
$i\sigma_y\otimes i\sigma_y\otimes I$ applied on particles $T$,
$A_1$ and $A_2$ can change the state \ket{\Psi_1} into
\ket{\Psi_7}. Trent compares his deduced result with the
authentication key they shared and then authenticates Alice$_1$
and Alice$_2$.

Now let us discuss the security for the present protocol. An
eavesdropper, Eve, has little chance to eavesdrop the users'
operation information because it is unnecessary for the users to
resend their particles on which each of users has performed their
corresponding operations according to their authentication keys,
to the trusted third party. Moreover, from the published results
of the users, Eve also cannot obtain any information of the users
because she has no Trent's result. Suppose the published results
of Alice$_1$ and Alice$_2$ are each \ket{\psi^-_{A_1A_1'}} and
\ket{\phi^+_{A_2A_2'}}. Without Trent's result, Eve can only know
that the state of the group is one of the four state
\{\ket{\Psi_5}$\otimes$\ket{\Psi_1},
\ket{\Psi_6}$\otimes$\ket{\Psi_1},
\ket{\Psi_7}$\otimes$\ket{\Psi_1},
\ket{\Psi_8}$\otimes$\ket{\Psi_1}\}. The eavesdropping check aims
to prevent Eve from impersonating attack and let the legal users
share a safe quantum channel with Trent. Suppose Eve prepares $N$
ordered three-particles GHZ states, each of which is
$\ket{\Psi_1}=\frac{1}{\sqrt{2}}(\ket{000}+\ket{111})_{FE_1E_2}$.
Eve intercepts particles $A_1$ and $A_2$ and resends particles
$E_1$ and $E_2$ to each Alice$_1$ and Alice$_2$. Eve attempts to
personate Trent for acquiring the users' authentication key.
However, during the eavesdropping check, Eve's attack will be
detected by the parties because Eve cannot tamper with the
classical message published by the trusted third party, Trent.
Thus the users' results have no correlation with the result
published by Trent.

According to Stinespring dilation theorem, Eve's action can be
realized by a unitary operation $\hat{E}$ on a large Hilbert
space, $H_{A_1A_2}\otimes H_{E}$. Then the state of Trent,
Alice$_1$, Alice$_1$ and Eve is
\begin{eqnarray}
\ket{\Phi}=\sum_{T,A_1,A_2\in\{0,1\}}\ket{\varepsilon_{T,A_1,A_2}}\ket{T}\ket{A_1A_2},
\end{eqnarray}
where \ket{\varepsilon} denotes Eve's probe state and \ket{T} and
\ket{A_1A_2} are states shared by Trent and the users. The
condition on the states of Eve's probe is
\begin{eqnarray}
\sum_{T,A_1,A_2\in\{0,1\}}\bra{\varepsilon_{T,A_1,A_2}}\;
\varepsilon_{T,A_1,A_2}\rangle=1.
\end{eqnarray}
As Eve can eavesdrop particle $A_1$ and $A_2$, Eve's action on the
system can be written as
\begin{eqnarray}
\label{security5}
\ket{\Phi}&=&\frac{1}{\sqrt{2}}[\ket{0}(\alpha_1\ket{00}\ket{\varepsilon_{000}}+\beta_1\ket{01}\ket{\varepsilon_{001}}+\gamma_1\ket{10}\ket{\varepsilon_{010}}\nonumber\\
&+&\delta_1\ket{11}\ket{\varepsilon_{011}})+\ket{1}(\delta_2\ket{11}\ket{\varepsilon_{100}}+\gamma_2\ket{10}\ket{\varepsilon_{101}}\nonumber\\
&+&\beta_2\ket{01}\ket{\varepsilon_{110}}+\alpha_2\ket{00}\ket{\varepsilon_{111}}].
\end{eqnarray}
The error rate introduced by Eve is
$\epsilon=1-|\alpha_1|^2=1-|\delta_2|^2$. Here the complex numbers
$\alpha$, $\beta$, $\gamma$ and $\delta$ must satisfy
$\hat{E}\hat{E}^\dag=I$.

We then generalize our three-party QIA protocol to a multiparty
one (more than three parties) (MQIA). In MQIA protocol, Trent can
authenticate many users, \{Alice$_1$, Alice$_2$, $\cdots$,
Alice$_r$\} ($r>2$) simultaneously. Trent prepares an ordered $N$
$(r+1)$-particle GHZ states
\begin{eqnarray}
\frac{1}{\sqrt{2}}(\ket{00\cdots0}+\ket{11\cdots1})_{T,A_1,\cdots,A_r}.
\end{eqnarray}
The details of MQIA is very similar to those of three-party one.
Trent sends $A_1$, $A_2$, $\cdots$, $A_r$ sequences to each
Alice$_1$, Alice$_2$, $\cdots$, Alice$_r$. Similar to step (S2),
Trent and the users check eavesdropping. If they confirm the
quantum channel is safe, they continue to the next step.
Otherwise, they abort the protocol. Trent divides the remaining
GHZ states into $M$ ordered groups, [\{P(1)$_{TA_1\cdots A_r}$,
Q(1)$_{T'A_1'\cdots A_r'}$\}, $\cdots$, \{P(M)$_{TA_1\cdots A_r}$,
Q(M)$_{T'A_1'\cdots A_r'}$\}]. Alice$_1$, Alice$_2$, $\cdots$,
Alice$_{(r-1)}$ each perform one of the two operations \{$I$,
$i\sigma_y$\} on their particles according to their authentication
keys. Trent then performs randomly $I$ or $i\sigma_y$ operation on
particle $T$ in each group. Each user measures particles $A_i$ and
$A'_i$ ($i=1,2,\cdots,r$) of each group in Bell basis. After
measurements, Alice$_1$, Alice$_2$, $\cdots$, Alice$_r$ publish
their measurement results. Trent performs Bell basis measurement
on particles $T$ and $T'$ of each group and authenticates the
users according to their measurement results.

In summary, we have presented a multiparty simultaneous quantum
identity authentication protocol based on entanglement swapping.
The trusted third party can authenticate many users
simultaneously. If there are many users waiting for being
authenticated by the system, the efficiency for identity
authentication can be improved greatly.



\begin{acknowledgments}
This work is supported by the National Natural Science Foundation of
China under Grant No. 60472032.
\end{acknowledgments}

%
%

%
%

\begin{thebibliography}{99}
\bibitem{bb84} C. H. Bennett and G. Brassard, \textit{in Proceedings of IEEE
international Conference on Computers, Systems and signal
Processing, Bangalore, India} (IEEE, New York), pp. 175 - 179
(1984).
\bibitem{hbb99} M. Hillery, V. Buz\v{e}k, and A. Berthiaume, Phys. Rev. A \textbf{59}, 1829 (1999).
\bibitem{kki99} A. Karlsson, M. Koashi, and N. Imoto, Phys. Rev. A \textbf{59}, 162 (1999).
\bibitem{zhang} Z. J. Zhang, Phys. Lett. A \textbf{342}, 60 (2005).
\bibitem{gg03} G. P. Guo and G. C. Guo, Phys. Lett. A \textbf{310}, 247 (2003).
\bibitem{zlm05} Z. J. Zhang, Y. Li, and Z. X. Man, Phys. Rev. A \textbf{71}, 044301 (2005).
\bibitem{xldp04} L. Xiao, G. L. Long, F. G. Deng and J. W. Pan, Phys. Rev. A \textbf{69}, 052307 (2004)
\bibitem{beige} A. Beige, B.-G. Englert, Ch. Kurtsiefer and H. Weinfurter, Acta Phys. Pol. A \textbf{101}, 357 (2002).
\bibitem{Bostrom} K. Bostr\"{o}em and T. Felbinger, Phys. Rev. Lett. \textbf{89}, 187902 (2002).
\bibitem{Deng} F. G. Deng, G. L. Long and X. S. Liu, Phys. Rev. A \textbf{68}, 042317 (2003).
\bibitem{denglong} F. G. Deng and G. L. Long, Phys. Rev. A \textbf{69}, 052319 (2004).
\bibitem{cai1} Q. Y. Cai and B. W. Li, Chin. Phys. Lett. \textbf{21}, 601 (2004).
\bibitem{cai4} Q. Y. Cai and B. W. Li, Phys. Rew. A \textbf{69}, 054301 (2004).
\bibitem{jwang1} J. Wang, Q. Zhang and C. J. Tang, quant-ph/0511092.
\bibitem{jwang2} J. Wang, Q. Zhang and C. J. Tang, quant-ph/0602166.
\bibitem{jwang3} J. Wang, Q. Zhang and C. J. Tang, quant-ph//0603100.
\bibitem{cw1} C. Wang, F. G. Deng, Y. S. Li, X. S. Liu and G. L. Long, Phys. Rev. A \textbf{71}, 044305 (2005).
\bibitem{cw2} C. Wang, F. G. Deng and G. L. Long, Opt. Commun. \textbf{253}, 15 (2005).
\bibitem{tg} T. Gao, F. L. Yan and Z. X. Wang, quant-ph/0406083.
\bibitem{zjz} Z. J. Zhang and Z. X. Man, quant-ph/040321.
\bibitem{gao1} T. Gao , F. L. Yan and Z. X. Wang, J. Phys. A \textbf{38}, 5761 (2005).
\bibitem{dhh99} M. Du$\check{s}$ek, O. Haderka and M. Hendrych, Phys. Rev. A \textbf{60}, 149 (1999).
\bibitem{zz00} G. H. Zeng and W. P. Zhang, Phys. Rev. A \textbf{61}, 022303 (2000).
\bibitem{m02} T. Mihara, Phys. Rev. A \textbf{65}, 052326 (2002).
\bibitem{hlee} H. Lee, J. Lim and H. Yang, Phys. Rev. A \textbf{73}, 042305 (2006).
\bibitem{zzhe93} M. Zukowski, A. Zeilinger, M. A. Horne and A. K. Ekert, Phys. Rev. Lett. \textbf{71}, 4287 (1993).
\end{thebibliography}
\end{document}